\documentclass[12pt]{iopart}

\usepackage{cite}
\usepackage{graphicx}
\usepackage{multirow}

\begin{document}

\title{Direct numerical method for counting statistics in stochastic processes}

\author{Jun Ohkubo and Thomas Eggel}

\address{
Institute for Solid State Physics, University of Tokyo, 
Kashiwanoha 5-1-5, Kashiwa, Chiba 277-8581, Japan
}
\ead{ohkubo@issp.u-tokyo.ac.jp}
\begin{abstract}
We propose a direct numerical method to calculate the statistics of the number of transitions
in stochastic processes, without having to resort to Monte Carlo calculations.
The method is based on a generating function method,
and arbitrary moments of the probability distribution of the number of transitions
are in principle calculated by solving numerically a system of coupled differential equations.
As an example, a two state model with a time-dependent transition matrix is considered
and the first, second and third moments of the current are calculated.
This calculation scheme is applicable for any stochastic process with a finite state space,
and it would be helpful to study current statistics in nonequilibrium systems.
\end{abstract}

\maketitle

\section{Introduction}

The statistics of nonequilibrium currents have attracted the interest of many physicists
(for example, see \cite{Derrida2007}.)
The statistics are related to the counting of  the number of specific transitions in stochastic processes,
and these problems have been widely studied in physics and chemistry
\cite{Bicout1999,Gopich2003,Zheng2003,Gopich2005,Sung2005,Gopich2006,Shikerman2008}.
The general scheme for counting the number of transitions, i.e., counting statistics,
has been recently developed (e.g., \cite{Gopich2005}),
and various exact and approximate analytical results have been found.
For example, a stochastic system under periodic perturbations
can exhibit a net current, a so-called pump current,
and it has been shown that this pump current is related to the geometric phase 
\cite{Sinitsyn2007,Ohkubo2008,Ohkubo2008a}.
For the adiabatic case in which the change of the periodic perturbations are very slow,
an analysis with the aid of the Berry phase is valid, 
and analytical expressions for the current (the first moment) and variance
have been obtained for a specific case \cite{Sinitsyn2007}.
However, to confirm these analytical expressions,
Monte Carlo calculations have been performed;
precisely speaking, although the first moment can be calculated
from solutions of the master equation,
Monte Carlo calculations were needed to estimate the variance numerically \cite{Sinitsyn2007}.
Although numerical simulations are helpful in order to study such time-dependent systems,
Monte Carlo calculations can become very costly, as discussed in \cite{Sinitsyn2007}.

In the present paper,
we derive a direct numerical method to calculate moments of the probability distribution
of the number of specific transitions in a stochastic process.
Although a numerical method to calculate photon emission statistics has been proposed recently
\cite{Peng2009},
the method has only been applied to quantum systems.
We will discuss in the following,
with the aid of the generating function approach developed by Gopich and Szabo \cite{Gopich2005},
a numerical method that applies to classical stochastic processes and in which the
numerical effort to obtain the moments of the probability distribution reduces to 
integrating numerically a system of coupled differential equations. 
That is, this numerical method does not involve Monte Carlo type calculations.
In addition, the method is applicable to a system under arbitrary time-dependent perturbations.
It would be difficult in general to obtain analytic expressions for current statistics
under complicated time-dependent perturbations,
so that the direct numerical method will be useful 
for investigations of current statistics in nonequilibrium systems.

The outline of the present paper is as follows.
In section 2 we explain the general scheme for the generating function approach.
Section 3 is the main part of the paper
and the direct numerical method is presented by way of an example in the pump current problem.
In addition, the validity of the proposed method
is confirmed by comparison with the Monte Carlo simulations in section 3.
Section 4 gives concluding remarks.

\section{General formalism for the generating function}

We first explain the general formalism for the generating function of the counting statistics.
Although the formalism is similar to the one proposed in \cite{Gopich2005},
we here present the general case in which the transition matrix can depend on time.

Denoting the probability of finding the system in state $n$ by $p_n(t)$,
a master equation for the system is
\begin{eqnarray}
\frac{\partial}{\partial t} p_n(t) = \sum_{m} \kappa_{nm}(t) p_m(t),
\end{eqnarray}
where $\{\kappa_{nm}(t)\}$ is a transition matrix.
The component $(n,m)$ of the transition matrix $\kappa_{nm}(t)$ is the rate of transition 
$m \to n$, and it can be time-dependent.

We here derive the generating function for counting the number of events 
of a specific transition $i_\mathrm{A} \to j_\mathrm{A}$.
The generalization to multiple transitions is straightforward.

Firstly, we denote the probability, with which the system starts in state $m$ and finishes in state $n$
with $N_\mathrm{A}$ being the number of transitions $i_\mathrm{A} \to j_\mathrm{A}$ during time $t$, 
as $P_{nm}(N_\mathrm{A}|t)$.
The probability $P_{nm}(N_\mathrm{A}|t)$ is obtained by repeated convolutions
of the probability of no transitions, i.e. 
\begin{eqnarray}
\fl
P_{nm}(N_\mathrm{A} | t ) = G_{n j_\mathrm{A}}'(t) \ast 
\underbrace{\kappa_{j_\mathrm{A} i_\mathrm{A}}(t) G_{i_\mathrm{A} j_\mathrm{A}}'(t) \cdots
\ast \kappa_{j_\mathrm{A} i_\mathrm{A}}(t) G_{i_\mathrm{A} j_\mathrm{A}}' (t)
}_{N_\mathrm{A} - 1}
\ast \kappa_{j_\mathrm{A} i_\mathrm{A}}(t) G_{i_\mathrm{A} m}' (t),
\end{eqnarray}
where $g_1(t) \ast g_2(t) \equiv \int_0^t g_1(t-t') g_2(t') \rmd t'$ denotes the convolution,
and $G_{kl}'(t)$ is the probability with which the system evolves from state $l$ to state $k$, 
provided no $i_\mathrm{A} \to j_\mathrm{A}$ transitions occur during time $t$.

Secondly, the generating function of the probability $P_{nm}(N_\mathrm{A}|t)$
is defined by
\begin{eqnarray}
f_{nm}(\lambda,t) = \sum_{N_\mathrm{A}}^\infty \lambda^{N_\mathrm{A}} P_{nm}(N_\mathrm{A}|t).
\end{eqnarray}
One can see that the generating function $f_{nm}(\lambda,t)$ 
satisfies the following integral equation
\begin{eqnarray}
f_{nm}(\lambda,t) = G_{nm}'(t) +
\int_0^t G_{n j_\mathrm{A}}' (t-t') \lambda \kappa_{j_\mathrm{A} i_\mathrm{A}}(t')
f_{i_\mathrm{A} m} (\lambda,t') \rmd t'.
\end{eqnarray}

Thirdly, we derive the time-evolution equation for the generating function
$f_{nm}(\lambda,t)$.
We notice that the probability of no transitions, $G_{nm}'(t)$, obeys
\begin{eqnarray}
\frac{\partial}{\partial t} G_{nm}'(t) = \sum_i \kappa_{ni} (t) G_{im}'(t)
- \delta_{n, j_\mathrm{A}} \kappa_{j_\mathrm{A} i_\mathrm{A}} (t) G_{i_\mathrm{A} m}'(t),
\end{eqnarray}
where $\delta_{i,j}$ is the Kronecker delta.
Hence, using the differentiation of the convolution,
\begin{eqnarray}
\frac{\partial}{\partial t} \int_0^t g_1(t-t') g_2(t') \rmd t'
= g_1(0) g_2(t) + \int_0^t  \left( \frac{\partial}{\partial t} g_1(t - t') \right) g_2(t') \rmd t',
\end{eqnarray}
the time-evolution equation for $f_{nm}(\lambda,t)$ is derived as follows:
\begin{eqnarray}
\fl
\frac{\partial}{\partial t} f_{nm} (\lambda,t) 
&= \sum_i \kappa_{ni}(t) G_{im}'(t) 
- \delta_{n, j_\mathrm{A}} \kappa_{j_\mathrm{A} i_\mathrm{A}}(t) G_{i_\mathrm{A} m}'(t)
+ \lambda G_{n j_\mathrm{A}}'(0) \kappa_{j_\mathrm{A} i_\mathrm{A}}(t)  f_{i_\mathrm{A} m}(t)
\nonumber \\
\fl
& \quad + \int_0^t \left( \frac{\partial}{\partial t} G_{n j_\mathrm{A}}'(t-t')\right)
\lambda \kappa_{j_\mathrm{A} i_\mathrm{A}}(t') f_{i_\mathrm{A} m}(t') dt' \nonumber \\
\fl
&=
\sum_i \kappa_{ni}(t) f_{im}(\lambda,t) - \delta_{n, j_\mathrm{A}} (1-\lambda) 
\kappa_{j_\mathrm{A} i_\mathrm{A}}(t) f_{i_\mathrm{A} m}(\lambda,t),
\label{eq_time_evolution_for_fnm}
\end{eqnarray}
where we used $G_{nm}'(0) = \delta_{n,m}$.
Equation \eref{eq_time_evolution_for_fnm} should be solved with the initial conditions
$f_{nm}(\lambda,0) = \delta_{n,m}$.

Finally, we construct the generating function for counting the number of the target transitions 
$i_\mathrm{A} \to j_\mathrm{A}$, i.e.,
\begin{eqnarray}
F(\lambda,t) = \sum_{N_\mathrm{A}=0}^\infty \lambda^{N_\mathrm{A}} P(N_\mathrm{A}|t).
\end{eqnarray}
Since $f_{nm}(\lambda,t)$ is the generating function with specific initial state $m$ and final state $n$,
the generating function with specific final state $n$ is constructed as
\begin{eqnarray}
\phi_n(\lambda,t) = \sum_m f_{nm}(\lambda,t) p_m(0),
\end{eqnarray}
where $p_m(0)$ is a probability distribution at initial time $t=0$.
Hence the generating function $F(\lambda,t)$ is calculated as
\begin{eqnarray}
F(\lambda,t) = \sum_n \phi_n(t).
\end{eqnarray}
In addition, the time evolution equation for the generating function $\phi_n(\lambda,t)$ is given by:
\begin{eqnarray}
\fl
\frac{\partial}{\partial t} \phi_{n}(\lambda,t)
&= \sum_{m} \frac{\partial}{\partial t} f_{nm}(\lambda,t) p_m(0) \nonumber \\
\fl
&= \sum_{m} \left[ 
\sum_{i} \kappa_{ni}(t) f_{im}(\lambda,t) - \delta_{n,j_\mathrm{A}} (1-\lambda)
\kappa_{j_\mathrm{A}i_\mathrm{A}}(t) 
f_{i_\mathrm{A}m}(\lambda,t)
\right] p_m(0) \nonumber \\
\fl
&= \sum_{i} \kappa_{ni}(t) \phi_{i}(\lambda,t) - \delta_{n,j_\mathrm{A}} (1-\lambda) 
\kappa_{j_\mathrm{A}i_\mathrm{A}}(t)
\phi_{i_\mathrm{A}}(\lambda,t),
\label{eq_time_evolution_for_phi}
\end{eqnarray}
and these equations should be solved with initial conditions 
$\phi_n(0) = \sum_m f_{nm}(\lambda,0) p_m(0)  = p_n(0)$.

\section{Direct numerical method for the moments}

\begin{figure}
\begin{center}
\includegraphics[width=80mm]{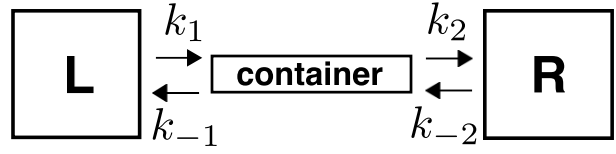}
\end{center}
\caption{A simple stochastic model with two states.} 
\label{fig_model}
\end{figure}

In order to explain the new direct numerical method to calculate moments in counting statistics,
we use a simple stochastic model with two states;
the discussion is easily extended to general cases with $N$ states.

The stochastic process with two states has been proposed 
in \cite{Sinitsyn2007} in the context of the pump current problem.
The system consists of three parts, as shown in Fig.~\ref{fig_model}.
The container can contain at most one particle.
When the container is filled with one particle
the particle can escape from the container by jumping into either one of the two particle baths,
i.e., the left reservoir [L] or the right one [R].
In the pump current problem, the transition rates $k_{\pm 1}$ and $k_{\pm 2}$ depend on time, 
and here we set them as
\begin{eqnarray}
\begin{array}{l}
k_{-1} = k_2 = 1, \\
k_1 = 1 + R \cos (\omega t), \\
k_{-2} = 1 + R \sin (\omega t).
\end{array}
\end{eqnarray}
Defining $p_1$ and $p_2$ as the respective probabilities 
of the container being empty or filled,
the master equation is written as
\begin{eqnarray}
\frac{\partial}{\partial t}
\left( \begin{array}{c} p_1(t) \\ p_2(t) \end{array} \right)
= 
\left(
\begin{array}{cc}
-k_1 - k_{-2} & k_{-1} + k_2  \\
k_1 + k_{-2}  & - k_{-1} - k_{2}
\end{array}
\right)
\left( \begin{array}{c} p_1(t) \\ p_2(t) \end{array} \right).
\label{eq_original_master_equation}
\end{eqnarray}

In order to estimate the particle current from the container to the right reservoir [R] we
need to count the number of transitions from the container to the right reservoir [R]
and the number of transitions from [R] to the container;
we should calculate the difference between them.
According to section~2, the generating function for the particle current
is given by
\begin{eqnarray}
F(\lambda,t) = \phi_1(\lambda,t) + \phi_2 (\lambda,t),
\end{eqnarray}
where $\phi_1(\lambda,t)$ and $\phi_2 (\lambda,t)$ obey the following time-evolution equations:
\begin{eqnarray}
&\frac{\partial}{\partial t} \phi_1(\lambda,t)
= (-k_1-k_{-2}) \phi_1 (\lambda,t) + (k_{-1}+k_2 \lambda) \phi_2 (\lambda,t), 
\label{eq_generating_1} \\
&\frac{\partial}{\partial t} \phi_2 (\lambda,t)
= (k_1+k_{-2} \lambda^{-1}) \phi_1 (\lambda,t) + (-k_{-1}-k_2) \phi_2 (\lambda,t).
\label{eq_generating_2}
\end{eqnarray}
Note that the transition from the container to [R] gives positive contributions
to the statistics, so that we multiply $k_2$ by $\lambda$.
In contrast, because the transition from [R] to the container gives negative contributions,
$\lambda^{-1}$ is multiplied to $k_{-2}$.

If we obtain the explicit solution of the generating function $\phi_i(\lambda,t)$,
all statistics of the particle current is estimated.
However, in general it is difficult to obtain the explicit solution of the generating function.
Instead of seeking analytic solutions, we proceed to obtain some moments directly via numerical calculations.
To this effect we need to derive the time-evolution equations of the moments.
Firstly, $\lambda$ in \eref{eq_generating_1} and \eref{eq_generating_2} is set to $1$:
\begin{eqnarray}
&\frac{\partial}{\partial t} \phi_1 |_{\lambda=1}
= (-k_1-k_{-2}) \phi_1|_{\lambda=1} + (k_{-1}+k_2) \phi_2|_{\lambda=1},  
\label{eq_generating_0th_order_1} \\
&\frac{\partial}{\partial t} \phi_2 |_{\lambda=1}
= (k_1+k_{-2}) \phi_1|_{\lambda=1} + (-k_{-1}-k_2) \phi_2|_{\lambda=1},
\label{eq_generating_0th_order_2}
\end{eqnarray}
where we denoted $\phi_i(\lambda=1,t)$ as $\phi_i |_{\lambda=1}$ for simplicity.
Because \eref{eq_generating_0th_order_1} and \eref{eq_generating_0th_order_2}
are exactly the same as the original master equation \eref{eq_original_master_equation},
$\phi_i |_{\lambda=1}$ is interpreted as the probability
in the original master equation, i.e.,
$\phi_1|_{\lambda=1} = p_1(t)$ and $\phi_2|_{\lambda=1} = p_2(t)$.
Next, the first derivatives of \eref{eq_generating_1} and \eref{eq_generating_2} with respect to $\lambda$
are calculated and again $\lambda$ is set to $1$:
\begin{eqnarray}
\fl
&\frac{\partial}{\partial t} \left. \frac{\partial \phi_1}{\partial \lambda} \right|_{\lambda=1}
= (-k_1-k_{-2}) \left. \frac{\partial \phi_1}{\partial \lambda} \right|_{\lambda=1}
+ k_2 p_2
+ (k_{-1}+k_2) \left. \frac{\partial \phi_2}{\partial \lambda} \right|_{\lambda=1},
\label{eq_generating_1st_order_1}
\\
\fl
&\frac{\partial}{\partial t} \left. \frac{\partial \phi_2}{\partial \lambda} \right|_{\lambda=1}
= - k_{-2} p_1
+  (k_1+k_{-2}) \left. \frac{\partial \phi_1}{\partial \lambda} \right|_{\lambda=1}
+ (-k_{-1}-k_2) \left. \frac{\partial \phi_2}{\partial \lambda} \right|_{\lambda=1}.
\label{eq_generating_1st_order_2}
\end{eqnarray}
Hence, the first moment of the particle current is calculated as follows:
\begin{eqnarray}
\frac{\partial}{\partial t} 
\langle n \rangle 
=
\frac{\partial}{\partial t} 
\left[ 
\left. \frac{\partial \phi_1}{\partial \lambda} \right|_{\lambda=1}
+ \left. \frac{\partial \phi_2}{\partial \lambda} \right|_{\lambda=1}
\right]
= k_2 p_2 - k_{-2} p_1,
\end{eqnarray}
which recovers the well-known result \cite{Sinitsyn2007}.

\begin{figure}
\begin{center}
\includegraphics[width=75mm]{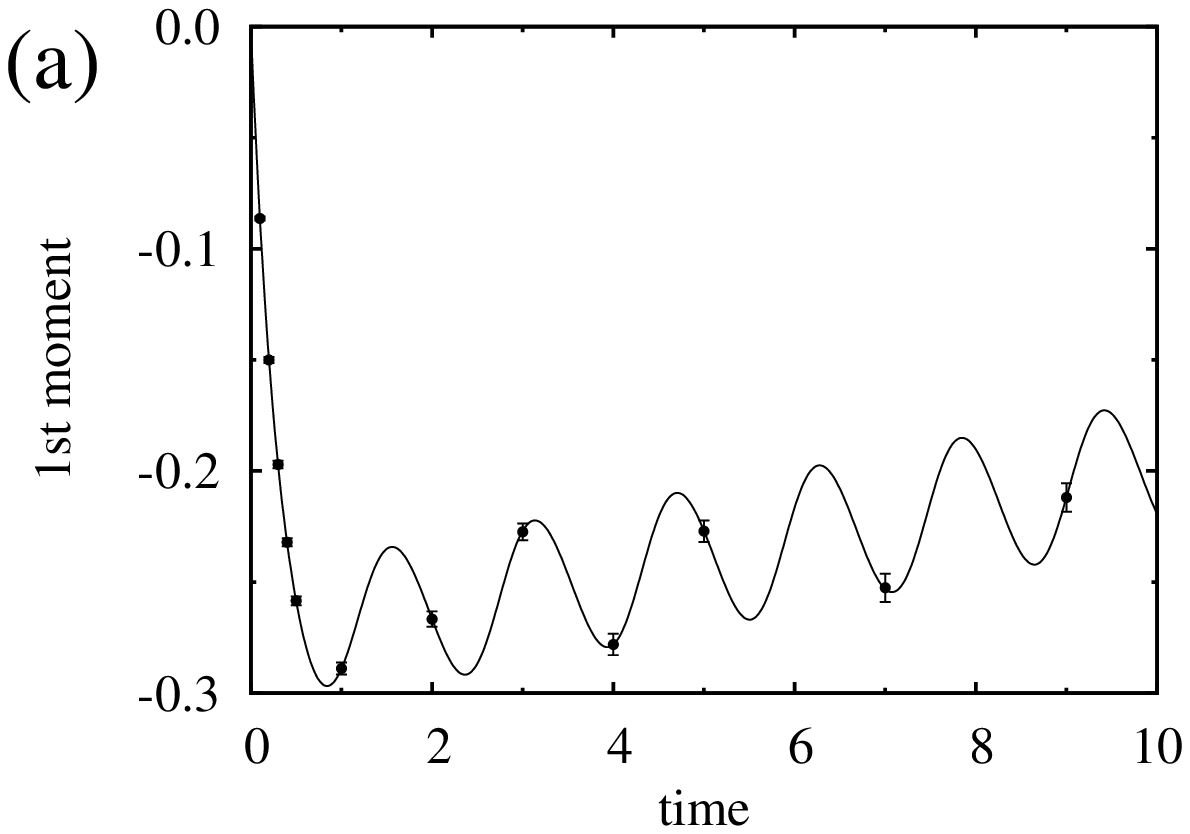}
\includegraphics[width=75mm]{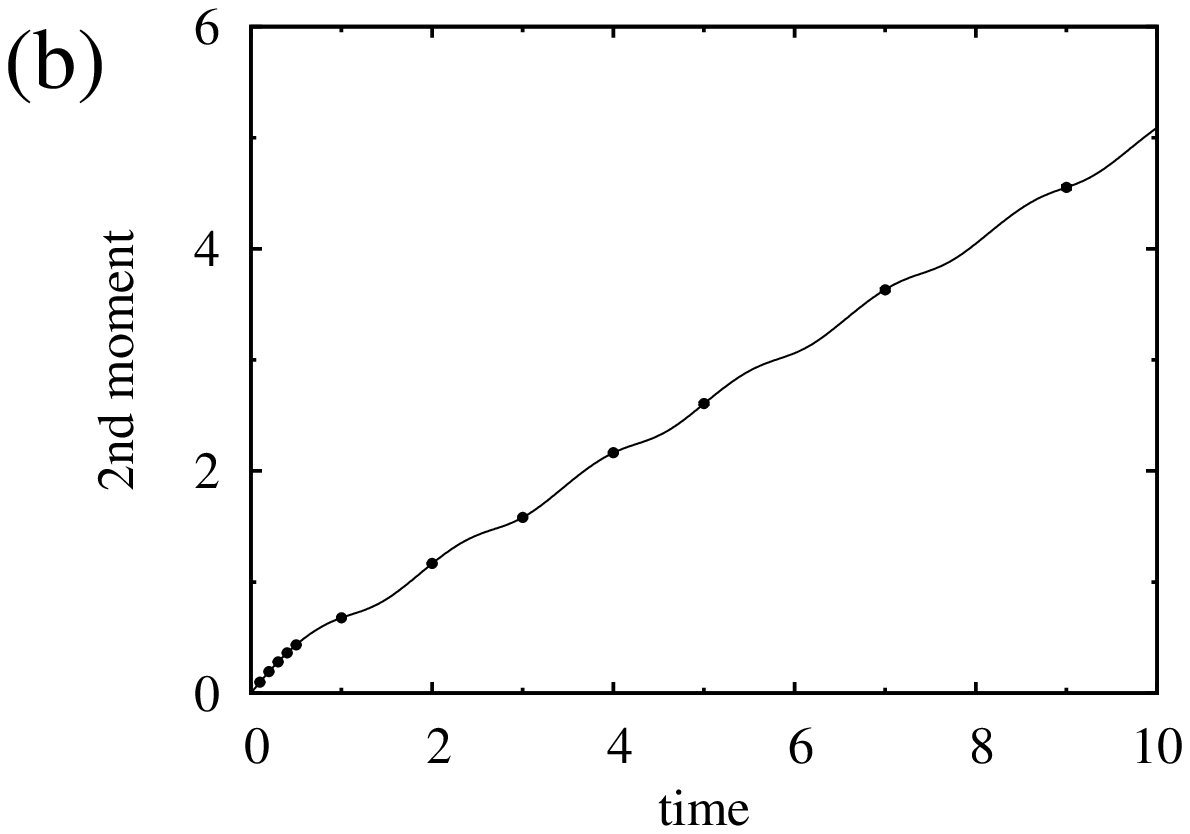}
\includegraphics[width=75mm]{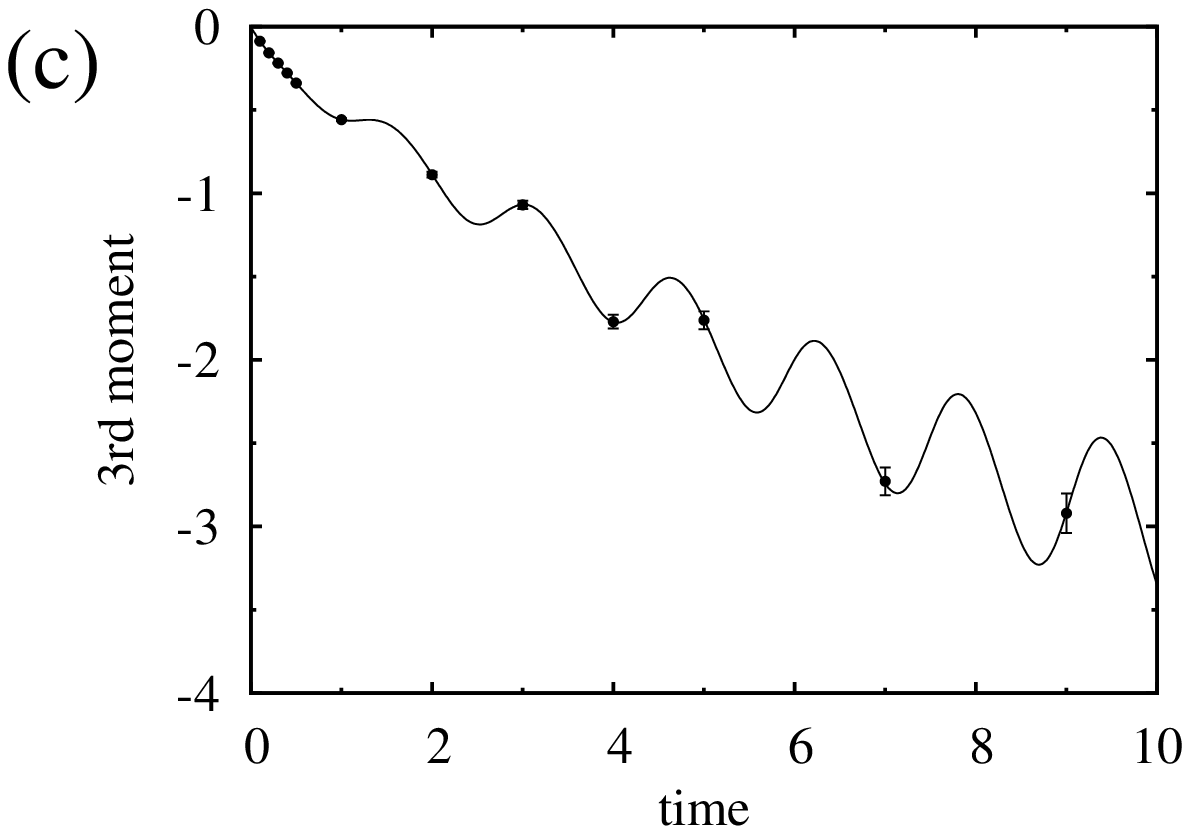}
\end{center}
\caption{Moments obtained by the new direct numerical method (solid lines) and
the Monte Carlo method (filled circles with error bars).
(a) first moment. (b) second moment. (c) third moment.}
\label{fig_results}
\end{figure}

The second and third `factorial' moments are also estimated as follows:
\begin{eqnarray}
\fl
\frac{\partial}{\partial t} 
\langle n (n-1) \rangle 
=
\frac{\partial}{\partial t} 
\left[ 
\left. \frac{\partial^2 \phi_1}{\partial \lambda^2} \right|_{\lambda=1}
+ \left. \frac{\partial^2 \phi_2}{\partial \lambda^2} \right|_{\lambda=1}
\right]
= 2 k_2 \left. \frac{\partial \phi_2}{\partial \lambda} \right|_{\lambda=1}
+ 2 k_{-2} p_1
- 2 k_{-2} \left. \frac{\partial \phi_1}{\partial \lambda} \right|_{\lambda=1}, \nonumber \\
\end{eqnarray}
and
\begin{eqnarray}
\fl
\frac{\partial}{\partial t} 
\langle n (n-1)(n-2) \rangle 
&= \frac{\partial}{\partial t} 
\left[ 
\left. \frac{\partial^3 \phi_1}{\partial \lambda^3} \right|_{\lambda=1}
+ \left. \frac{\partial^3 \phi_2}{\partial \lambda^3} \right|_{\lambda=1}
\right] \nonumber \\
\fl
&= 3 k_2 \left. \frac{\partial^2 \phi_2}{\partial \lambda^2} \right|_{\lambda=1}
- 6 k_{-2} p_1
+ 6 k_{-2} \left. \frac{\partial \phi_1}{\partial \lambda} \right|_{\lambda=1}
- 3 k_{-2} \left. \frac{\partial^2 \phi_1}{\partial \lambda^2} \right|_{\lambda=1}.
\end{eqnarray}
The time evolution equations for $\partial^2 \phi_i / \partial \lambda^2$
are easily obtained, in a similar way to \eref{eq_generating_1st_order_1} 
and \eref{eq_generating_1st_order_2}.
In numerical estimation of the time development of the differential equations, 
all $\partial^n \phi_i / \partial \lambda^n$ are set to $0$ initially,
because $\phi_i(\lambda,0) = p_i(0)$ does not depend on $\lambda$ initially.
It is easy to obtain the second and third moments from the above factorial moments.

In order to calculate the $m$-th moment,
we should solve $m \times 2$ coupled differential equations.
Generally, when a stochastic process consists of $N$ states,
we need $m \times N$ coupled differential equations
to obtain moments up to $m$-th order.
Hence, without using any Monte Carlo method,
we can deterministically calculate the moments for the particle current 
or the number of specific transitions.

To confirm the validity of the new direct numerical method,
we compare results obtained from the new method with those of the Monte Carlo method,
i.e., the time-dependent Gillespie algorithm \cite{Anderson2007}.
The parameters are $R = 0.5$ and $\omega = 4.0$,
and the initial state is $p_1 = 1$ (hence, $p_2 = 0$);
the container is empty at the initial time.
In one data set, $10^5$ Monte Carlo trajectories were used to estimate moments of the particle current.
We repeated the Monte Carlo calculations, and collected $100$ data sets.
The error bars in Fig.~\ref{fig_results} correspond to 
the standard deviation in the $100$ data sets.
Although we can calculate moments at arbitrary time in the Monte Carlo calculations,
we plotted only some points for reference.
The results of the new method and those of the Monte Carlo method agree completely.

\section{Concluding remarks}

In the present paper, 
a new numerical method to estimate moments for current statistics was proposed.
The proposed method needs only time integrations
of a system of coupled differential equations,
so that the moments are estimated deterministically without the aid of Monte Carlo simulations.
By way of an example,
we explained the proposed method using a simple two-state model,
but it is easy to apply this method to an arbitrary stochastic process with a finite state space.

In studies of a nonequilibrium current or nonequilibrium properties,
it would be valuable to obtain detailed information about the statistics of the current.
If one can obtain analytical solutions for the generating function,
all statistics, including the probability distribution for the current,
are calculated from the solutions.
However, it is difficult in general to treat a case with a complicated time-dependent transition matrix.
On the other hand, the proposed numerical method is available to arbitrary time-dependency,
and we believe that the proposed method is both important and useful in cases in which an analytical solution of the nonequilibrium current is out of reach.

\section*{Acknowledgments}
This work was supported in part by grant-in-aid for scientific research 
(Grants No.~20115009 and No.~21740283)
from the Ministry of Education, Culture, Sports, Science and Technology (MEXT), Japan.
T.E. is supported by a Government Scholarship from the MEXT.

\end{document}